\documentclass[runningheads,a4paper]{llncs}

\usepackage{amssymb}
\usepackage{graphicx}
\usepackage{epstopdf}
\usepackage{subfig}
\usepackage{multirow}
\usepackage{amsmath}
\usepackage{array}
\usepackage{calc}
\usepackage{url}
\usepackage[export]{adjustbox}
\usepackage[belowskip=-15pt,aboveskip=0pt]{caption}
\let\llncssubparagraph\subparagraph
\let\subparagraph\paragraph
\usepackage[compact]{titlesec}
\let\subparagraph\llncssubparagraph
\newcolumntype{C}[1]{>{\centering\let\newline\\\arraybackslash\hspace{0pt}}m{#1}}

\urldef{\mailsa}\path|{alfred.hofmann, ursula.barth, ingrid.haas, frank.holzwarth,|
\urldef{\mailsb}\path|anna.kramer, leonie.kunz, christine.reiss, nicole.sator,|
\urldef{\mailsc}\path|erika.siebert-cole, peter.strasser, lncs}@springer.com|
\newcommand{\keywords}[1]{\par\addvspace\baselineskip
\noindent\keywordname\enspace\ignorespaces#1}

\begin{document}
\parskip 0pt
\mainmatter  

\title{Formal Dependability Modeling and Analysis: A Survey\thanks{ The final publication is available at http://link.springer.com}}

\titlerunning{Formal Dependability Modeling and Analysis: A Survey}

%
%
\author{Waqar Ahmed \inst{1} %
\and Osman Hasan \inst{1} \and Sofi\`ene Tahar \inst{2}}
%

\institute{School of Electrical Engineering and Computer Science\\
National University of Sciences and Technology,
Islamabad, Pakistan\\
\email{ \{waqar.ahmad,osman.hasan\}@seecs.nust.edu.pk  }
\and Electrical and Computer Engineering Department\\
Concordia University, Montreal, Canada \\
\email{tahar@ece.concordia.ca}
}

%
%

\maketitle

\begin{abstract}
\label{abst}
Dependability is an umbrella concept that subsumes many key properties about a system, including reliability, maintainability, safety, availability, confidentiality, and integrity. Various dependability modeling techniques have been developed to effectively capture the failure characteristics of systems over time. Traditionally, dependability models are analyzed using paper-and-pencil proof methods and computer based simulation tools but their results cannot be trusted due to their inherent inaccuracy limitations. The recent developments in probabilistic analysis support using formal methods have enabled the possibility of accurate and rigorous dependability analysis. Thus, the usage of formal methods for dependability analysis is widely advocated for safety-critical domains, such as transportation, aerospace and health. Given the complementary strengths of mainstream formal methods, like theorem proving and model checking, and the variety of dependability models judging the most suitable formal technique for a given dependability model is not a straightforward task. In this paper, we present a comprehensive review of existing formal dependability analysis techniques along with their pros and cons for handling a particular dependability model.

\end{abstract}

\section{Introduction}
\label{sec:intro}
The rapid advancement in technology in the past few decades has enabled us to develop many sophisticated systems that range from ubiquitous hand-held devices (like cell phones and tablets) to high-end computing equipment used in aircrafts, power systems, nuclear plants and healthcare devices. Ensuring the reliable functioning of these sophisticated systems is a major concern for design engineers. This concern is greatly amplified for safety-critical systems where a slight malfunction in the system may endanger human lives or lead to heavy financial set-backs. In order to avoid such scenarios beforehand, several dependability modeling  techniques have been developed that can effectively model the failure characteristics of a system  and thus analyze its failure behavior.

\textit{Dependability} is primarily defined as the ability of a system to deliver
service that can justifiably be trusted \cite{avizienis2001fundamental}. Dependability is an umbrella concept which is evolved from \textit{reliability} and \textit{availability} considerations \cite{avizienis2001fundamental}. Many authors describe dependability of a system as a set attributes, such as reliability, maintainability, safety, availability, confidentiality, and integrity \cite{spitzer2000digital}. Some of these attributes, i.e. reliability and availability, are quantitative whereas some are qualitative, for instance, safety \cite{avizienis2001fundamental}.

\textit{Reliability} is defined as the probability of a system or a sub-component functioning correctly under certain conditions over a specified interval of time \cite{avizienis2001fundamental}. Availability is a  closely related concept to reliability and it can be defined as the probability that a component will be available when demanded \cite{avizienis2001fundamental}. To understand the difference between reliability and availability, it is important to realize that reliability refers to failure-free operation during an interval, while availability refers to failure-free operation at a given instant of time \cite{avizienis2001fundamental}. Availability can be viewed as a special case of reliability and is thus  commonly considered as an attribute of reliability \cite{al2009comparative}. The availability of a system is typically measured as a function of reliability and \textit{maintainability}, which is defined as the the probability of performing a successful repair action of a system under a given time and stated conditions \cite{avizienis2001fundamental}. Additionally, if we keep the maintainability measure constant, the availability of the system is directly proportional to the reliability of the system \cite{Weibull_15}. This paper mainly focuses on  reliability and availability attributes of dependability, since maintainability can be considered as a part of availability.

 The first step in conducting the dependability analysis is the calculation of basic metrics of reliability and availability, such as mean-time to failure (MTTF) \cite{avizienis2001fundamental}, mean-time between failure (MTBF) \cite{avizienis2001fundamental} and mean-time to repair (MTTR) \cite{avizienis2001fundamental}, at the individual \textit{component level} of the given system. The next step is the selection of an appropriate dependability modeling technique. Some of the widely used dependability modeling techniques include Reliability Block Diagrams (RBD) \cite{vcepin2011reliability}, Fault Trees (FT) \cite{vesely1981fault} and Markov chains (MC) \cite{gilks2005markov}. The selection among these modeling techniques depends upon numerous factors, which include the level of available details, size and complexity of the given communication network system. These modeling techniques allow us to estimate the reliability and availability of the system at the \textit{system level} and play a particularly useful role at the design stage of a system for scrutinizing the design alternatives without building the actual system. Once the modeling technique is selected, the third and the last step is the choice of the appropriate \textit{system level} reliability and availability analysis technique. The dependability models, formed using these techniques,  are analyzed using paper-and-pencil based analytical methods  or simulation. However, these analysis methods cannot ascertain absolute correctness of the analysis mainly because of the human error and manual manipulations involved in the former and the sampling based deduction and the usage of pseudo random numbers and computer arithmetic in the later.  Formal methods, on the other hand,  use mathematical logic to precisely model the system's intended behavior and deploy mathematical reasoning to construct an irrefutable proof that the given system satisfies its requirements. This kind of mathematical modeling and analysis makes formal methods an accurate and rigorous analysis method compared to the traditional analytical and simulation based analysis.  Thus, they are being strongly advocated for being used for the dependability analysis of safety-critical systems.

The purpose of this survey paper is to provide a generic overview of the formal methods that are being utilized  for dependability analysis. These formal methods primarily include: (i) Petri Nets (ii) Model Checking and (iii) Higher-order Logic theorem proving as they have all been used for the dependability analysis using the three dependability modeling techniques: RBD, FT, and MC. The main focus of the paper is to study the utilization of formal methods in conjunction with the dependability modeling techniques for real-world applications and thus gain insights about the strengths and weaknesses of these formal methods and how to use them in the most effective manner. It is important to note that the paper is unique compared to existing surveys and tutorials on dependability analysis \cite{trivedi1993reliability,bernardi2012dependability,venkatesan2013survey,al2009comparative} due to its exclusive focus on dependability modeling techniques and their analysis with formal methods. For instnace, in \cite{trivedi1993reliability} a unified framework for reliability with Markov reward models is described and then a survey of existing reliability analysis software tools is presented. Similarly, a  survey of work related to dependability modeling and analysis of software and systems specified with UML is presented in \cite{bernardi2012dependability}. In \cite{venkatesan2013survey} and \cite{al2009comparative}, tools and methods that have been used for enhancing the dependability of Wireless Sensor networks (WSN) and communication networks are also surveyed, respectively. Unlike above work, this paper discusses about the pros and cons of modeling techniques and formal methods for the dependability analysis of a broad range of systems.


The organization of the paper is as follows: Section 2  briefly describes commonly used dependability modeling techniques. Section 3 presents a detailed survey of formal methods that have been used for conducting accurate and rigorous dependability analysis of real-world systems. Section 4 provides the insights and the common pitfalls of the dependability modeling techniques and also a comparison of formal methods with traditional dependability analysis techniques. Finally, Section 5 concludes the paper.

\section{Dependability Modeling Techniques}
\label{sec:depend_model_tec}

Dependability assessment techniques can be utilized in every design phase of the system or component including development, operation and maintenance. FT and RBD based models are usually used to provide reliability and availability estimates for both \textit{early} and \textit{later} stages of the design, where the system models are more refined and have more detailed specifications compared to the early stage system models \cite{avizienis2001fundamental}. While on the other hand, MC based models are mainly used in the \textit{later} design phase to perform trade-off analysis among different design alternatives when the detailed specification of the design becomes available. In addition, when the system is deployed, these modeling techniques can be beneficial in order to estimate the frequency of maintenance and part replacement in the design, which allows us to determine the life cost of the system elements or components. In this section, we present a brief detail about some commonly used dependability modeling techniques to facilitate the understanding of the next sections.

\subsection{Reliability Block Diagrams}
\label{subsec:RBD}
Reliability Block Diagrams (RBD) \cite{trivedi2008probability} are graphical structures consisting of blocks and connector lines. The blocks usually represent the system components and the connection of these components is described by the connector lines.  The system is functional, if at least one path of properly functional components from input to output exists otherwise it fails.

\begin{table}[!ht]
	\centering
	\caption{RBDs with their Mathematical Expressions}
	\scriptsize
	\scalebox{0.9}{
		\begin{tabular}{|l |c|}
			\hline
			RBDs & Mathematical Expressions  \\
			\hline
			\hline
			
			\includegraphics[valign=c,scale=0.2,trim={0cm 0cm 0 0cm},clip,natwidth=610,natheight=642]{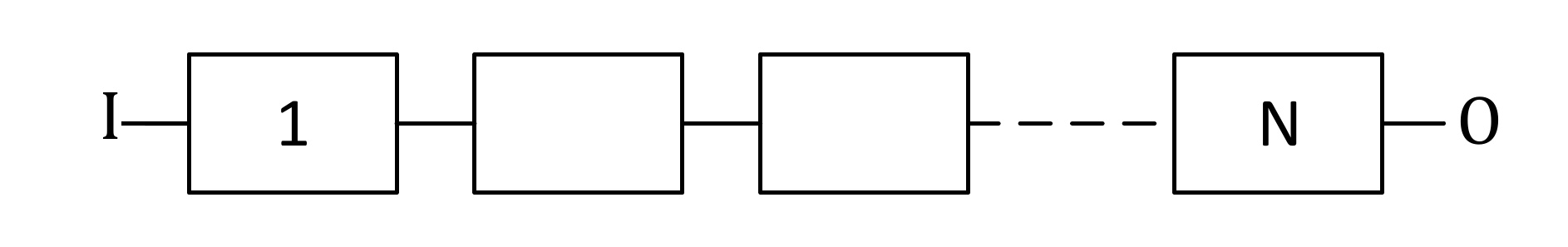} & 	$\!\begin{aligned}[t] &\small{
				R_{series}(t) = Pr (\bigcap_{i=1}^{N}A_{i}(t))
				= \prod_{i=1}^{N}R_{i}(t)
			}\end{aligned}$ \\
			\hline
			\includegraphics[valign=c,scale=0.2,trim={0cm 0cm 0 0cm},clip,natwidth=610,natheight=642]{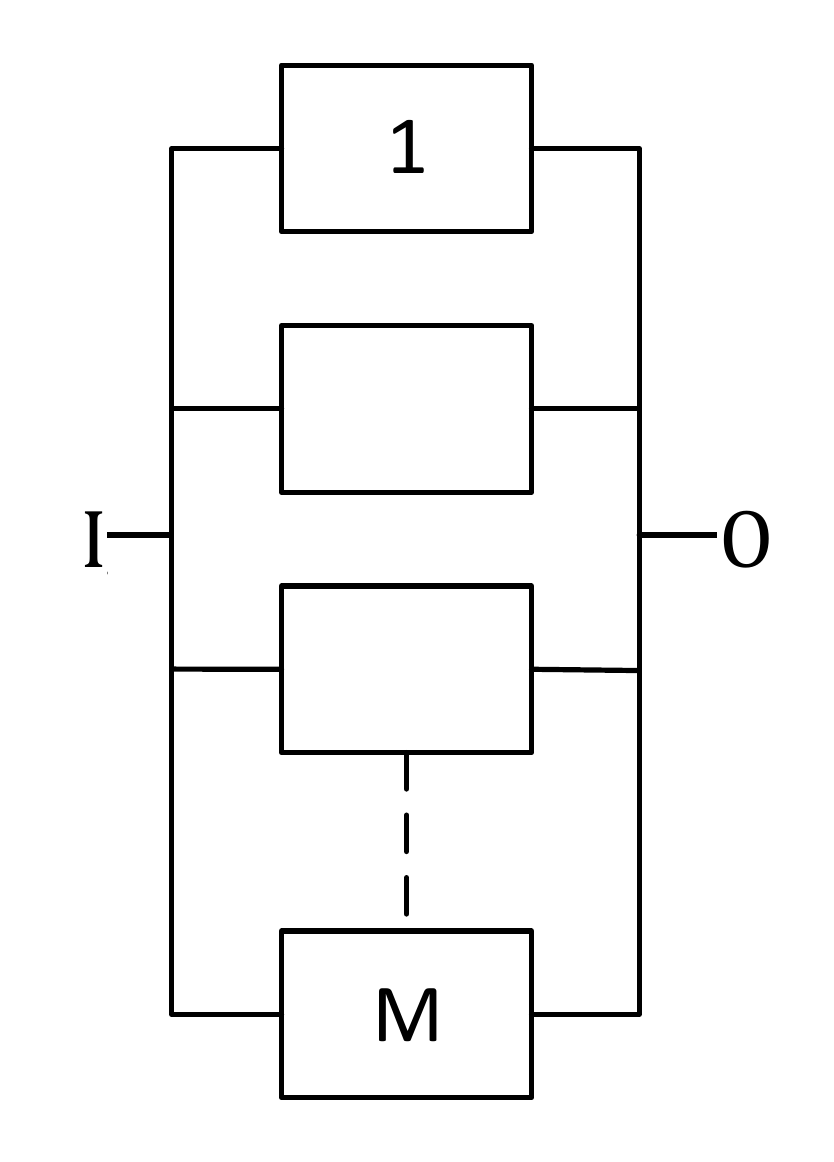}& $\!\begin{aligned}[c]\small{R_{parallel}(t) = Pr (\bigcup_{i=1}^{M}A_{i} ) = 1 - \prod_{i=1}^{M}(1 - R_{i}(t))
			}\end{aligned}$ \\
			\hline		
			\includegraphics[valign=c,scale=0.2,trim={0cm 0cm 0 0cm},clip,natwidth=610,natheight=642]{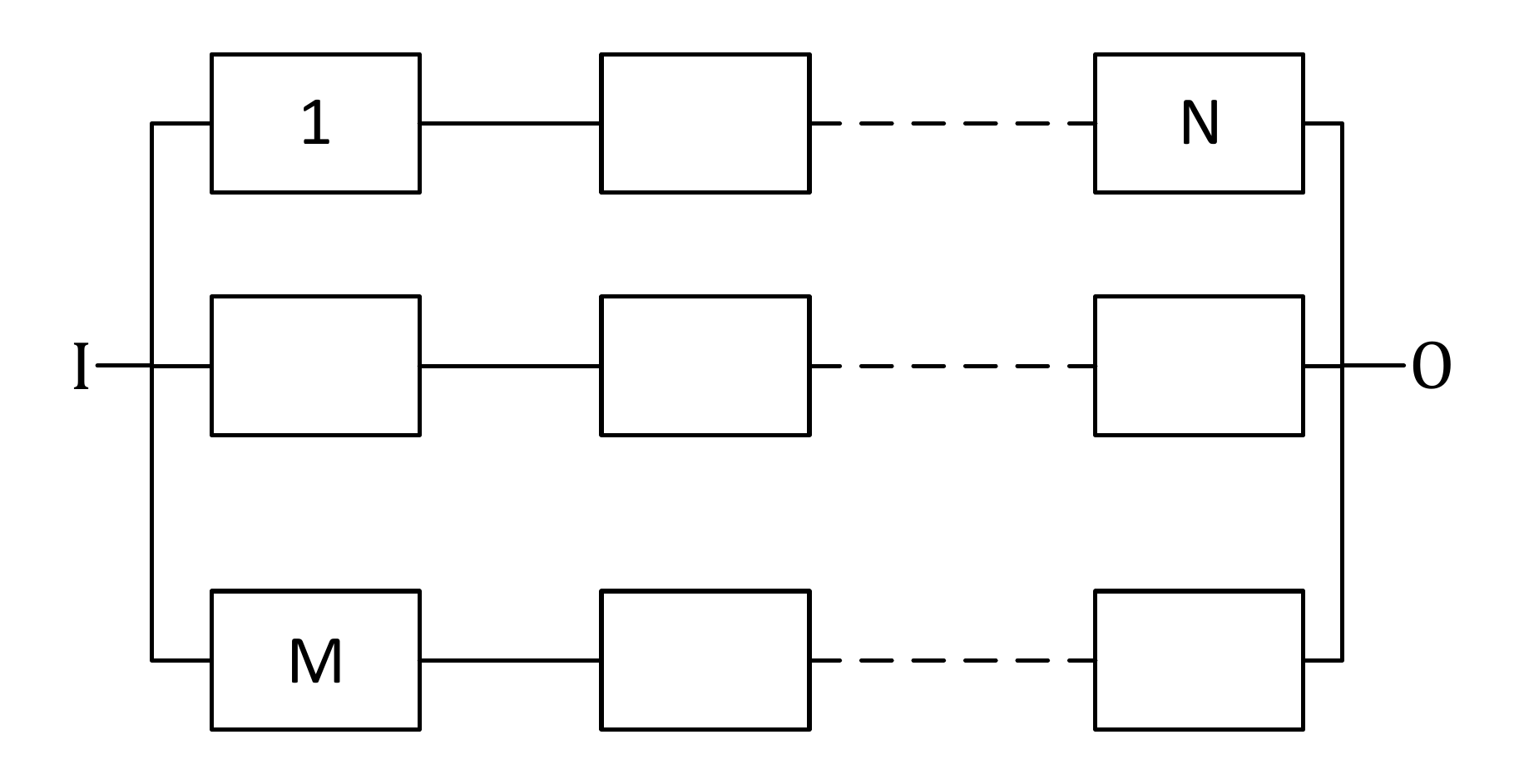} & $\!\begin{aligned}[c]  \small{R_{parallel-series}(t)} & \small{= Pr (\bigcup_{i=1}^{M} \bigcap_{j=1}^{N} A_{ij}(t))} \\ & \small{= 1- \prod_{i=1}^{M}(1 - \prod_{j=1}^{N} (R_{ij}(t)))}\end{aligned}$ \\
			\hline					
			\includegraphics[valign=c,scale=0.2,trim={0 0cm 0cm 0cm},clip,natwidth=610,natheight=642]{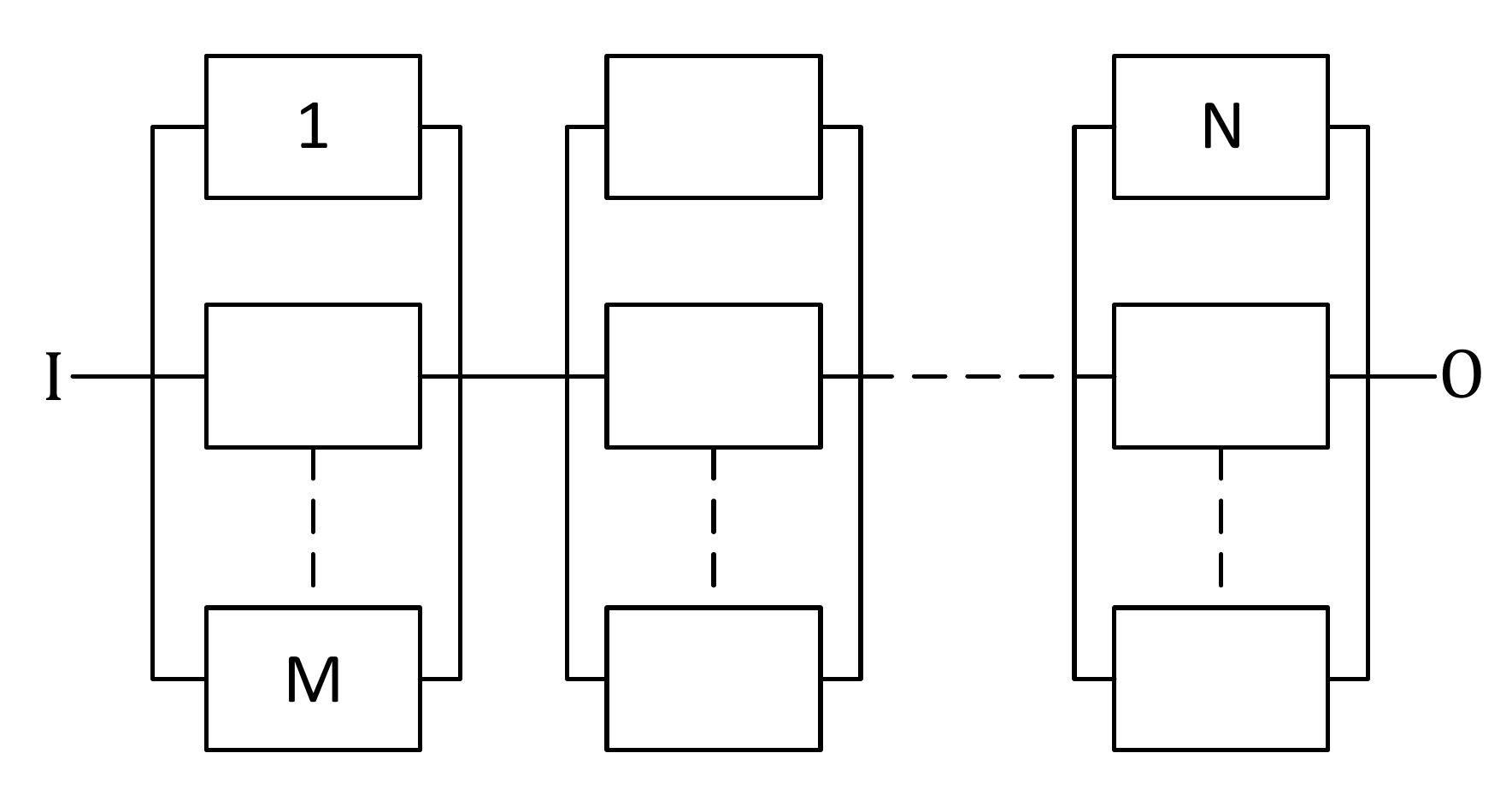} & $\!\begin{aligned}[c]  \small{R_{series-parallel}(t)} & \small{= Pr (\bigcap_{i=1}^{N} \bigcup_{j=1}^{M} A_{ij}(t))}\\ &  \small{= \prod_{i=1}^{N}(1 - \prod_{j=1}^{M} (1- R_{ij}(t)))}\end{aligned}$ \\
			\hline
\includegraphics[valign=c,scale=0.2,trim={0 0cm 0cm 0cm},clip,natwidth=610,natheight=642]{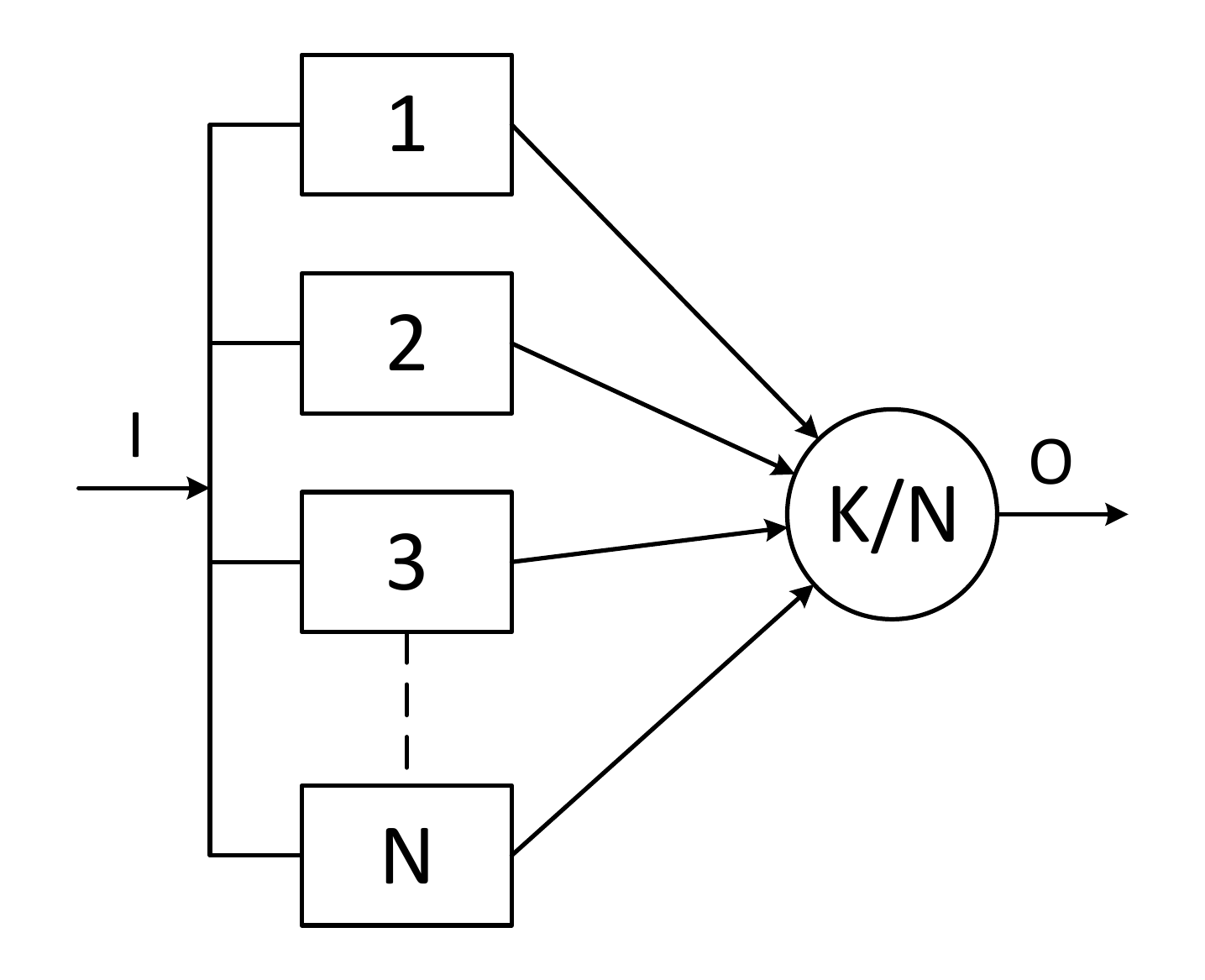} & $\!\begin{aligned}[c]  \small{R_{k|n}(t)} & \small{= Pr (\bigcup_{i=k}^{n}\{\textit{exactly \textit{i} components functioning}\})}\\ &  \small{= \Sigma_{i=k}^{n} (\dbinom{n}{k} R^{i} (1 - R)^{n -1})}\end{aligned}$ \\
			\hline
		\end{tabular}}\label{table:RBD_exp}
	\end{table}	

An RBD construction can follow any one of three basic patterns of component connections: (i) series (ii) active redundancy or (iii) standby redundancy. In the \textit{series} connection, shown in Table \ref{table:RBD_exp}, all  components should be functional for the system to remain functional. The corresponding reliability expression is also shown in Table \ref{table:RBD_exp}, where $A_{i}$ represents the event corresponding to $i^{th}$ component.  In an \textit{active} redundancy, all components in at least one of the redundant stages must be functioning in fully operational mode. The components in an active redundancy might be connected in a parallel structure or a combination of series and parallel structures as shown in Table \ref{table:RBD_exp}. In a \textit{standby} redundancy, all components are not required to be active. In other words, at least $k$ out of $n$ are required by the system to be functional, which can be seen in Table \ref{table:RBD_exp}.  There are three main requirements for building the RBD of a given system, i.e., the information about the (i) functional interaction of the system components; (ii) reliability of each component usually expressed in terms of failure distributions, such as exponential or Weibull, having appropriate failure rates; and (iii) mission times at which the reliability is desired. This information is then utilized by the design engineers to identify the appropriate RBD configuration (series, parallel or series-parallel) in order to determine the overall reliability of the given system. The detail about these commonly used RBD configurations and their corresponding mathematical expressions are presented in Table \ref{table:RBD_exp}.

\subsection{Fault Trees}
\label{subsec:FT}
Fault Tree (FT ) \cite{vesely1981fault} is a graphical technique for analyzing the conditions and the factors causing an undesired \textit{top event}, i.e., a critical event, which can cause the whole system failure upon its occurrence. These causes of system failure are represented in the form of a tree rooted by the \textit{top event}. The preceding nodes of the fault tree are represented by \textit{gates}, which are used to link two or more \textit{cause events} causing one fault in a prescribed manner. For example, an OR FT gate can be used when one fault suffices to enforce the fault. On the other hand, the AND FT gate is used when all the cause events are essential for enforcing the fault. Besides these gates, there are some other gates, such as exclusive OR FT gate, priority FT gate and inhibit FT gate, which can be used to model the occurrence of faults due to the corresponding cause events \cite{vesely1981fault}.

\begin{table}[t!]
	\centering
	\caption{Probability of Failure of Fault Tree Gates}
	\scriptsize
	\begin{tabular}{|l|l|}
		\hline
		FT Gates & Failure Probability Expressions \\
		\hline
		\hline
		\includegraphics[valign=c,scale=0.15,trim={0cm 0cm 0 0cm},clip,natwidth=610,natheight=642]{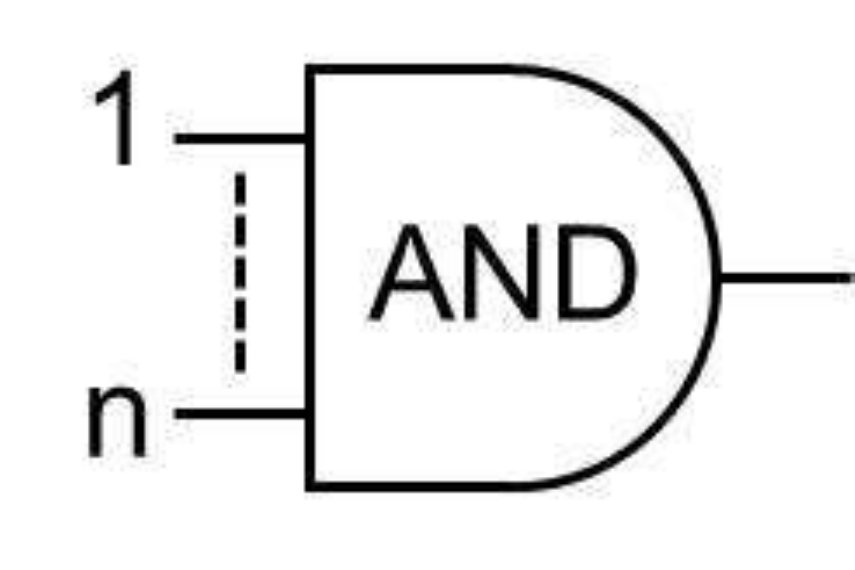}& $\!\begin{aligned}[t]
		\small{F(t)
			= Pr (\bigcap_{i=2}^{N}A_{i}(t))
			= \prod_{i=2}^{N}F_{i}(t)}
		\end{aligned}$
		\\
		\includegraphics[valign=c,scale=0.15,trim={0cm 0cm 0 0cm},clip,natwidth=610,natheight=642]{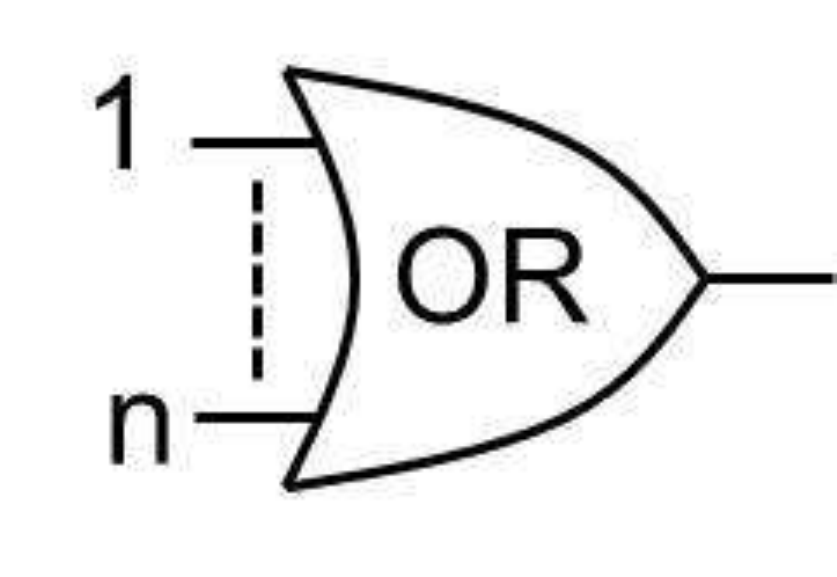}& $\!\begin{aligned}[t]
		\small{F(t)
			= Pr (\bigcup_{i=2}^{N}A_{i}(t))
			= 1 - \prod_{i=2}^{N}(1 - F_{i}(t))}
		\end{aligned}$
		\\
		\includegraphics[valign=c,scale=0.15,trim={0cm 0cm 0 0cm},clip,natwidth=610,natheight=642]{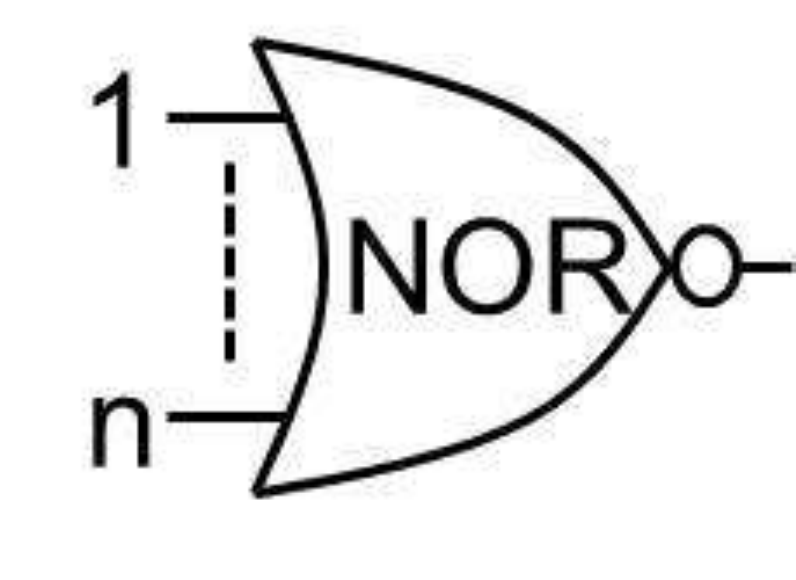}& $\!\begin{aligned}[t]
		\small{F(t)
			= 1 - F_{OR}(t) = \prod_{i=2}^{N}(1 - F_{i}(t))}
		\end{aligned}$
		\\
		\includegraphics[valign=c,scale=0.15,trim={0cm 0cm 0 0cm},clip,natwidth=610,natheight=642]{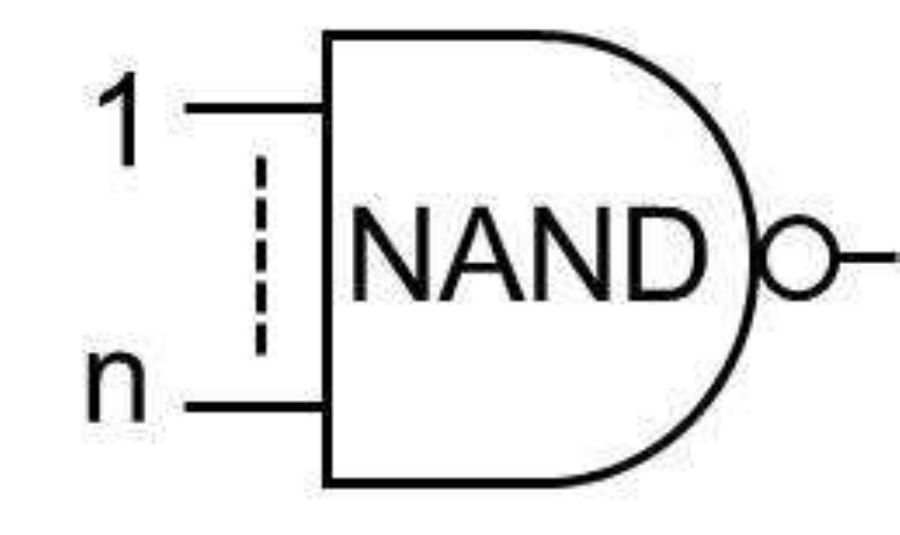}&  $\!\begin{aligned}[t]
		\small{F(t)} \small{=  Pr (\bigcap_{i=2}^{k}\overline A_{i}(t) \cap \bigcap_{j=k}^{N}A_{i}(t))} \small{= \prod_{i=2}^{k}(1 - F_{i}(t)) *\prod_{j=k}^{N}(F_{j}(t))}\end{aligned}$   \\
		\includegraphics[valign=c,scale=0.15,trim={0cm 0cm 0 0cm},clip,natwidth=610,natheight=642]{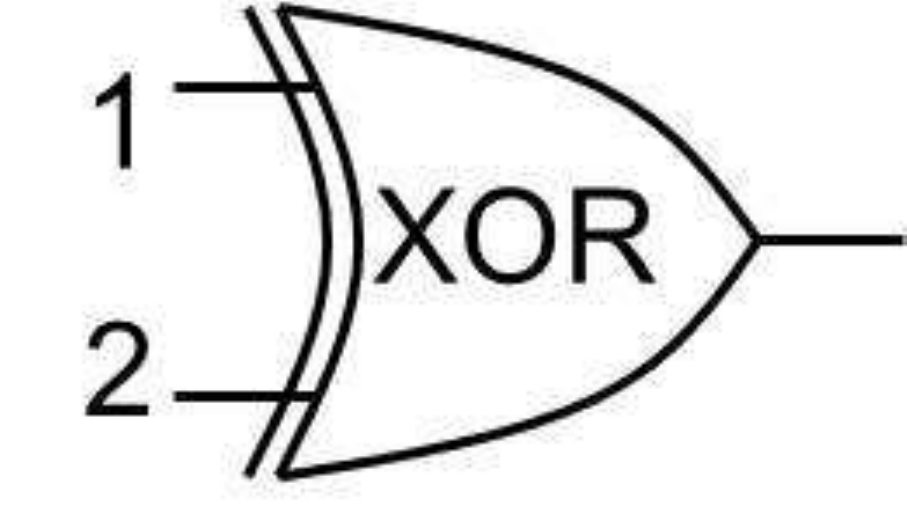}& $\!\begin{aligned}[t]
		\small{F(t)} \small{= Pr(\bar{A}(t)B(t) \cup A(t)\bar{B}(t))} \small{= F_{A}(t)(1 - F_{B}(t)) +  F_{B}(t)(1 - F_{A}(t))}\end{aligned}$    \\
		
		\includegraphics[valign=c,scale=0.15,trim={0cm 0cm 0 0cm},clip,natwidth=610,natheight=642]{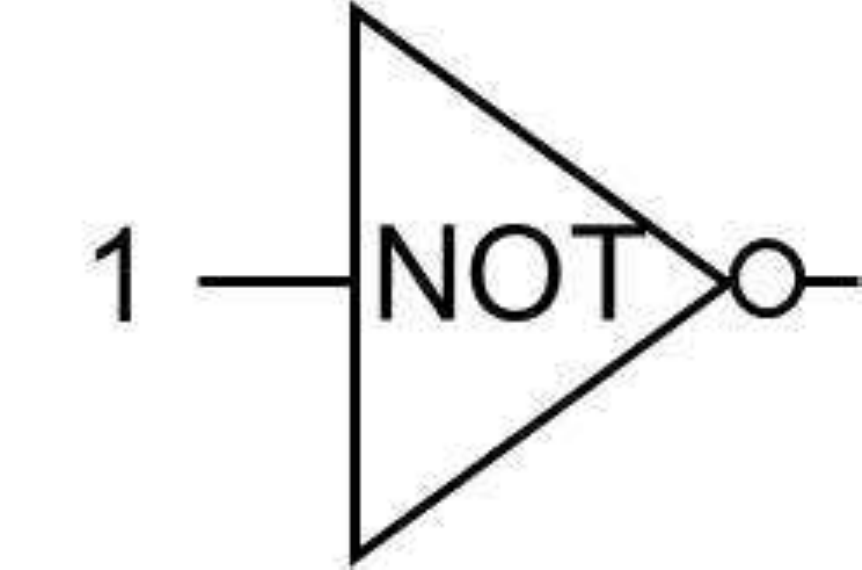}& $\!\begin{aligned}[t]
		F(t)&= Pr(\bar{A}(t))=(1 - F_{A}(t))\end{aligned}$     \\
		\hline
	\end{tabular}\label{FT_table}
\end{table}

Once the fault tree model is constructed, both qualitative and quantitative analysis can be carried out. A qualitative analysis in this context allows the identification of all combinations of basic failure events, known as cut sets, which can cause the top event to occur. The \textit{minimal} cut sets (MCS) are those cut sets that do not contain any subset of the basic cause events that are still a \textit{cut set} and are obtained by applying Boolean algebraic operations on these cut sets. The smaller the number of basic cause events in these cut set, the more resilient to failures is the considered modeled system. The quantitative analysis is used to evaluate the probability of occurrence of the top event by considering these minimal cut sets, which significantly contribute to the system failures.

In Fault Tree analysis (FTA), each FT gate has an associated failure probability expression as shown in Table \ref{FT_table}. These expressions can be utilized to evaluate the reliability of the system. The first step in the FTA is the construction of the FT of the given system. This is followed by the assignment of the failure distributions to basic $cause$-$events$ and the identification of the Minimal Cut Set (MCS) failure events, which contribute in the occurrence of the top event. These MCS failure events are generally modeled in terms of the \textit{exponential} or \textit{Weibull} random variables and the Probabilistic Inclusion-Exclusion (PIE) principle \cite{trivedi2008probability} is then used to evaluate the probability of failure of the given system.

\subsection{Markov Chain}
\label{subsec:MC}
A MC \cite{fugua2003applicability} is a stochastic process that consists of a set of states, i.e., $S =\{s_{0},s_{1},...,s_{n}\}$, and arcs, which are used to point the transition from one state to another. The initial state $s_{ini}$ and the probability $p_{ij}$ represent the starting state and the transition probability from state $s_{i}$ to state $s_{j}$, respectively. The process starts from an initial state and transitions from the current state to the next state occur on the basis of transition probabilities, which only depend upon the current state based on  the Markov or the memoryless property. Markov chains are usually classified into two categories: Discrete Time Markov Chains (DTMC) and Continuous Time Markov Chains (CTMC). Markovian models are frequently utilized for reliability analysis in scenarios where failure or repair events can occur at any point in time \cite{fugua2003applicability}.

Markov modeling has also been utilized for analyzing the \textit{dynamic} behavior of the other reliability models, i.e., RBD and FT. The notion of dynamic behavior, for reliability analysis, represents the evolution of system topology/configuration with respect to time. In the case of Dynamic Reliability Block Diagrams (DRBD) \cite{distefano2006new}, the system is modeled in terms of \textit{states} of the components and the evolution of these components states is carried out by a sequence of \textit{events} \cite{distefano2006new}. A typical DRBD contains the following states: (i) \textit{Active}: the state of proper functioning of the component; (ii) \textit{Failed}: the failure state of the component; and (iii) \textit{Standby}: the state depicting the case when the component is not in functional or in active condition but it can be activated. In addition, there are other states such as \textit{Hot}, \textit{Warm} and \textit{Cold}, representing the conditions when the system or component is disabled but energized, partially and completely disabled, respectively \cite{distefano2006new}.

\section{Formal Dependability Analysis Techniques}
\label{sec:form_depend_anal_tec}

\subsection{Petri Nets}

A Petri Net (PN) \cite{peterson1981Petri} is a bipartite directed graph consisting of disjoint sets of places $P$ and transitions $T$. The former, which is represented by circles, models the condition while the latter, signified by bars, represents the events or activities that may occur in the system. The directed arcs $(P \times T)$ and  $(T \times P)$, represented by arrows, describe the input places $P$ for the transitions $T$ and output places $P$ for the transitions $T$, respectively. Places may be empty or contain more than one token that is drawn by a block dot and term \emph{marking} represents the tokens over the set of places. A transition is said to be enabled, in a given marking, if all its input places contain at least one token. An enabled transition can be $fired$ and as a result a token will be removed from the input places of the transition and added to its output places.

Petri Nets and its variants are widely used as a reliability analysis tool for many real-world systems due
to their ability to efficiently handle large problems of dynamic
nature. For instance, PNs have been used for the reliability assessment of Web services \cite{zhong2006petri} and a wind turbine hydraulic variable pitch system \cite{yang2011petri}. Many existing work have utilized the PNs for \emph{availability} analysis, for instance, the availability of a mechanical system is hierarchically analyzed by dividing the complete system into three levels \cite{kumar2012reliability}. A system level PN model is constructed by composing the PNs of the subsystem levels, which are also composed from the PNs of the component level. Similarly, PNs have been used to analyze the availability of computational servers that are processing the jobs in a queue \cite{jian2008petri}, a replicated file system to reduce the overhead in a distributed environment \cite{dugan1989stochastic}, a subsea blowout preventer (BOP), which is essentially required to provide safety for drilling workers, rigs and natural environment \cite{zengkai2013availability} and the C160 series equipment that can modify its own modules based on different process plan and forms a new configuration \cite{beirong2012availability}. In addition, a considerable amount of work has been done by utilizing PN in conjunction with the dependability modeling techniques, described in Section \ref{sec:depend_model_tec}, for dependability analysis as follows:

\paragraph{\bf{Reliability Block Diagrams}}
 Many PN variants are extensively utilized to represent the RBDs to model the reliability of communication systems with dynamic nature.  For instance, the live migration process in cloud computing networks makes the system dynamic and thus yields to a complex RBD model, which can be effectively handled using Petri Nets with the support of commercial tools, such as \textit{SNOOPY} \cite{snoopy} and \textit{CPN} \cite{beaudouin2001cpn}. Given the dynamic nature of visualization, due to the presence of hardware systems, software systems, live migration techniques, resource allocation algorithms and concurrent failures, virtualized networks are frequently modeled with RBDs, which are then transformed to Petri Nets for the reliability analysis \cite{wei2011dependability}. The reliability of communication networks with \textit{redundancy mechanisms} has also been efficiently analysed using RBD based Petri Nets \cite{guimaraes2011dependability}.

  PNs have also been used to ensure the security/safety aspects of networks in terms of reliability and availability by analyzing the safety/security aspects of network protocols,  such as internet voting systems \cite{omidi2012modeling} and high-speed trains \cite{lijie2012verification}.
 In addition to the communication network, PNs have been used to develop the RBDs to analyze the reliability of a logistic supply chain \cite{li2014calculation} and redundant electrical generator used to power-up the coast guard vessel \cite{robidoux2010automated}. Similarly, a Cojoint system model consisting of CPN and RBD has been effectively used to analyze the dependability and logistics of a fault-redundant
 space station \cite{nebel2008modeling}.

\paragraph{\bf{Fault Trees}}
The PN approach has also been utilized, in conjunction with FTs, for the reliability analysis of embedded systems by translating the PN reachability into provability of linear logic sequents, which empowers the analysis by utilizing sequent calculus \cite{sadou2009reliability}. The \textit{dynamic behavior} of networks components, such as timed behavioral nature, cannot be captured by simple FT models but PNs provide a very feasible alternative for this purpose. The system under consideration is modeled with a FT, which is then transformed into its corresponding PN based model for analysis. For example, the reliability of the broadband integrated service network (B-ISDN) has been assessed by modeling the dynamic re-routing mechanism of the traffic using the FT-based PN approach \cite{balakrishnan1996stochastic}.

\paragraph{\bf{Markov Chains}}
A considerable amount of work has been done on analyzing reliability of systems using PNs with Markov chains.  Some other prominent work in this direction include the reliability analysis of a preemptive M/D/1/2/2 client-server queuing system \cite{radevanalysis}, the dynamic reconfiguration of FPGA \cite{kohlik2009dependability},  the data communication systems of the WLAN based train control system \cite{zhu2012service}, cellular networks \cite{jindal2011markov} and Wireless Sensor Networks (WSN) \cite{schoenen2013erlang}. Moreover, some network protocols, like the  courier  \cite{youness2006robust} and Fibre Distributed Data Interface (FDDI) token ring protocol \cite{christodoulou1994petri}, have also been analyzed using the Petri Net approach. Similarly, the reliability of a file server system \cite{ibe1993performance}, financial system \cite{279721}, distributed memories  \cite{sun2009survivability} and Low Earth Orbit (LEO) satellite has also been analyzed using PNs based on Markov chains \cite{zeng2011spn}. Moreover, a Markov regenerative PN has been introduced in \cite{choi1994markov} to extend the capability of stochastic PN analysis and then its effectiveness is illustrated by utilizing this approach to approximate client-server systems.

\subsection{Model Checking}
\label{subsec:model_check}
Model Checking \cite{baier2008principles} allows to describe the behavior of a given system in the form of a state machine and verify its temporal  properties in a rigorous manner. Probabilistic model checking extends traditional model checking principles for the analysis of MCs and allows the verification of probabilistic properties. Some notable probabilistic model checking include \textit{PRISM} \cite{lin2010modeling} and \textit{ETMCC} \cite{hermanns2003etmcc}.

Probabilistic model checking techniques have been considerably adopted to verify the reliability and availability properties of many systems, for instance, the \textit{PRISM} has been used to assess the reliability of e-health systems used in hospitals based on the Fast Health Interoperable Resources (FHIR) standard \cite{pervez2014formal} and the Device Interoperability Middleware (DIM) used to bridge the gap between different healthcare vendors \cite{pervez2015formal}. In addition, the \textit{PRISM} model checker has been utilized for the reliability/safety analysis of airbone applications by augmenting it to the Matlab simulink \cite{gomes2010systematic}, a RAID disk protocol used for reading the data from the disk sectors \cite{gopinath2009reliability}, multi-processor systems based on the Triple modular redundancy (TMR) model \cite{ge2010analysing}. \textit{PRISM} has also been utilized for quantitative reliability and availability analysis of a satellite system \cite{peng2013probabilistic}.

\paragraph{\bf{Fault Trees}} The \textit{COMPASS} tool \cite{bozzano2009compass} supports the formal FT analysis, specifically for aerospace systems. For verification purposes, \textit{COMPASS} provides support of several model checking tools, like \textit{NuSMV} \cite{cimatti2002nusmv} and \textit{MRMC} \cite{katoen2005Markov}. This tool provide various templates containing placeholders that have to be filled in by the user. These templates are primarily composed of the most frequently used patterns that allow easy specifications of properties by non-experts by hiding the details of the underlying temporal logic. The tool generates several outputs, such as traces, FTs and Failure Mode and Effect Analysis (FMEA) tables, along with diagnostic and performance measures.

 \paragraph{\bf{Markov Chains}} Probabilistic model checking extends traditional model checking principles for the analysis of MCs and allows the verification of probabilistic properties. Probabilistic model checking techniques have been considerably adopted to verify the reliability properties of many systems, such as NAND multiplexing \cite{norman2005evaluating}, an airbag system, an industrial process control system and the Herschel-Planck satellite system \cite{norman2014quantitative}. In \cite{conghua2013analysis}, the reliability analysis of the Fast And Secure Protocol (FASP) is carried out by first defining the successful data transmission using STL and then the communication network is modeled in the form of a sender, receiver and a communication channel module in \textit{PRISM}. Finally, the reliability property is then verified against the communication network using the \textit{PRISM} model checker.

 \subsection{Higher-order-Logic Theorem Proving}
 \label{subsec:HOL}
  Interactive theorem provers, like \textit{HOL4}, \textit{Isabelle/HOL} and \textit{Coq}, can be used to reason about probabilistic behaviors using the higher-order-logic formalizations of probability theory \cite{hurd_02,mhamdi_lebsegue,holzl2011three}.  This feature has been widely used to conduct the dependability analysis of many systems. For instance, the probability theory in  \textit{HOL4} \cite{mhamdi_lebsegue} has been used for the reliabililty analysis of combinational circuits \cite{hasan2011formal} and
 reconfigurable memory arrays \cite{hasan2010formal}. In these work, however, the reliability is evaluated based on probabilistic principles directly, i.e., no component to system-level assessment based on RBD or FT methods is done.  Similarly, formally verified statistical properties of the continuous  random variables have been used to reason about the fundamental reliability properties, including survival function and hazard rate \cite{abbasi2010formal}. These reliability properties are then used to analyzed the reliability of electronic system components \cite{abbasi2010formal}.

\paragraph{\bf{Reliability Block Diagrams}} The higher-order logic theorem prover \textit{HOL4} has been recently used for the formalization of RBDs, including series \cite{WAhmad_CICM14}, parallel
\cite{WAhmed_Wimob15}, parallel-series \cite{WAhmed_Wimob15} and series-parallel \cite{WAhmed_IWIL15}. These formalizations have been used for the
reliability analysis of a simple oil and gas pipeline with serial components \cite{WAhmad_CICM14}, WSN protocols \cite{WAhmed_Wimob15}
and logistic supply chains \cite{WAhmed_Wimob15}.

\paragraph{\bf{Fault Trees}} A higher-order-logic formalization of generic Fault Tree gates, i.e., AND, OR, NAND, NOR, XOR and NOT and the formal verification of their failure probability expressions  have also been  recently proposed in \textit{HOL4} \cite{WAhmad_CICM15}. In addition, this work also presents a formalization of  probabilistic inclusion-exclusion principle, which is then used to conduct the FT-based failure analysis of a solar array used in a Dong Fang Hong-3 (DFH-3) satellite  \cite{WAhmad_CICM15}.

\paragraph{\bf{Markov Chains}} A foundational formalization of time-homogeneous DTMC with finite state space has been presented in HOL4 \cite{liu2013formal} and \textit{Isabelle/HOL}
 \cite{holzl2012interactive}. These formalizations have been successfully used to formally analyze a binary communication channel \cite{liu2013formal}, ZeroConf \cite{holzl2012interactive} and anonymizing crowds protocols \cite{holzl2012interactive}.  None of these Markov chain formalizations has been used for reliability analysis so far.

\section{Comparison and Discussion}
\label{sec:InsightsPitfalls}

\subsection{Comparison of Dependability Modeling Techniques}
The criteria for the selection of these modeling techniques, for a certain system, mainly depends upon the type of system and problem domain. A  comparison among these modeling techniques is
shown in Table \ref{RFM_comp}. For instance, RBD is primarily used if we are interested in the successful working of the system while FT models the failure relationship due to the failure of individual components of the system. Also, both of these techniques utilize top-down analysis approach that starts at the system level and then proceeds downward to link system performance to failures at the component level. Due to this reason, these techniques work only for combinatorial types of problems, where a combination of components faults is used to determine the overall system failure. On the other hand, Markov chains are more flexible in terms of handling a wide variety of problems, as given in Table \ref{RFM_comp}, including non-combinatorial problems, where systems are in different operational modes, such as active or failed. However, Markov chains fail to cater for large and complex systems due to the exponential growth in the number of states.

\begin{table}[!htb]
	\caption{Comparison of Dependability Modeling Techniques}
	\centering
	\small
		\begin{tabular}{|p{0.4\linewidth}|p{0.2\linewidth}|p{0.1\linewidth}|p{0.1\linewidth}|}
			\hline
			Features & Reliability Block Diagram & Fault Tree & Markov Chain\\
			\hline
			Success Domain & $\checkmark$  &  & $\checkmark$\\
			\hline
			Failure Domain &  & $\checkmark$  & $\checkmark$ \\
			\hline
			Top-Down Approach & $\checkmark$  & $\checkmark$  & $\checkmark$\\
			\hline
			Identification and Prevention of Faults& $\checkmark$  & $\checkmark$  & $\checkmark$\\
			\hline
			Combinatorial Problems& $\checkmark$  & $\checkmark$ & $\checkmark$ \\
			\hline
			Non-combinatorial Problems &  &  & $\checkmark$ \\
			\hline
			Large and Complex Systems & $\checkmark$  & $\checkmark$ & \\
			\hline
		\end{tabular}\label{RFM_comp}
	\end{table}
	
	Based on the survey conducted in Section \ref{sec:form_depend_anal_tec}, we have found that FTs have been the mostly utilized dependability modeling technique by formal methods. On the other hand, the utilization of RBD and MC models for the dependability analysis is rapidly increasing specifically by PNs. The usage of RBD models with model checking for the formal dependability analysis is an area that is almost unexplored. We believe that this combination of modeling and analysis technique has a huge potential for ensuring accurate reliability analysis of a wide variety of safety-critical system.
	
	\subsection{Comparison of Dependability Analysis Techniques}
	
	A summary of  various dependability analysis techniques  is presented in Table \ref{table:comparison_tech}. These techniques are evaluated according to their expressiveness, accuracy and the possibility of automating the analysis. Model checking and Petri Nets are not expressive enough to model and verify all sorts of reliability properties due to their state-based nature. The accuracy of the paper-and-pencil based proofs is questionable because they are prone to human errors. Simulation is inaccurate due to the involvement of pseudo-random number generators and computer arithmetics along with its inherent sampling-based nature. Theorem proving does not support all the reliability analysis foundations as of now. Finally, the paper-and-pencil based proof methods and interactive theorem proving based analysis involve human guidance and therefore are not categorized as automatic. However, there is some automatic verification support (e.g. \cite{slind2008brief}) available for theorem proving, which can ease the human interaction in proofs and thus we cannot consider interactive theorem proving as a completely manual approach. All three formal methods techniques promise to provide accurate results and thus can be very useful for analyzing the dependability aspects of safety and financial-critical systems.
	
	 	\begin{table}[!h]\caption{Comparison of Reliability Analysis Techniques}
	 		\centering
	 		\scriptsize
	 			\begin{tabular}{|p{2.5cm}|p{2cm}|p{1.5cm}|p{1.5cm}|p{1.5cm}|p{1.5cm}|}
	 				\hline
	 				Feature & Paper-and-pencil Proof & Simulation Tools& Petri Nets &Theorem Proving &Model Checking \\
	 				\hline
	 				Expressiveness & $\checkmark$ & $\checkmark$& & $\checkmark$ &  \\
	 				\hline
	 				Accuracy & $\checkmark$  (?)&  & $\checkmark$ & $\checkmark$ & $\checkmark$ \\
	 				\hline
	 				Automation &  & $\checkmark$ & $\checkmark$ &   & $\checkmark$\\
	 				\hline
	 			\end{tabular}\label{table:comparison_tech}
	 		\end{table}
We have used the question mark symbol in accuracy feature  for paper-and-pencil to highlight its limitation of being prone to human error.

	\section{Conclusions}
	\label{sec:conclusions}	
	In this paper, we have discussed various dependability models constructed using the building blocks offered by the formalisms of reliability block diagrams, fault trees and Markov chains models. We have also presented a critical comparison, of the various dependability analysis techniques, i.e., analytical methods, simulation, and formal methods. Apart from providing the necessary background, we have also provided a detailed survey of the application of formal methods available in the open literature focused on studying dependability analysis of various real-world systems. The main contribution of this paper is that it is the first work presenting a comprehensive review of the various dependability modeling techniques in conjunction with formal methods along with a critical analysis describing their pros and cons in various contexts. Existing surveys on dependability analysis are either focused on software or communications networks and do not cover formal methods in depth.	\section*{Acknowledgments}
	This publication was made possible by NPRP grant \# [5 - 813 - 1 134] from the Qatar National Research Fund (a member of Qatar Foundation). The statements made herein are solely the responsibility of the author[s].
		
\small
\bibliographystyle{splncs}
\footnotesize{
\bibliography{formal}}

\begin{thebibliography}{10}

\bibitem{avizienis2001fundamental}
Avizienis, A., Laprie, J.C., Randell, B.:
\newblock {Fundamental Concepts of Dependability}.
\newblock Technical Report CS-TR-739, Newcastle University, UK (2001)
  \url{http://pld.ttu.ee/IAF0530/16/avi1.pdf}.

\bibitem{spitzer2000digital}
Spitzer, C.R., Spitzer, C.:
\newblock {Digital Avionics Handbook}.
\newblock CRC Press (2000)

\bibitem{al2009comparative}
Al-Kuwaiti, M., Kyriakopoulos, N., Hussein, S.:
\newblock {A Comparative Analysis of Network Dependability, Fault-tolerance,
  Reliability, Security, and Survivability}.
\newblock Communications Surveys \& Tutorials \textbf{11}(2) (2009)  106--124

\bibitem{Weibull_15}
Weibull:
\newblock http://www.weibull.com/hotwire/issue26/relbasics26.htm (2015)

\bibitem{vcepin2011reliability}
{\v{C}}epin, M.:
\newblock {Reliability Block Diagram}.
\newblock In: Assessment of Power System Reliability.
\newblock Springer (2011)  119--123

\bibitem{vesely1981fault}
Vesely, W.E., Goldberg, F.F., Roberts, N.H., Haasl, D.F.:
\newblock {Fault tree handbook (NUREG-0492)}.
\newblock Technical report, U.S. Nuclear Regulatory Commission (1981)

\bibitem{gilks2005markov}
Gilks, W.R.:
\newblock {Markov chain Monte Carlo}.
\newblock Wiley Online Library (2005)

\bibitem{trivedi1993reliability}
Trivedi, K.S., Malhotra, M.:
\newblock {Reliability and performability techniques and tools: A survey}.
\newblock In: Messung, Modellierung und Bewertung von Rechen-und
  Kommunikationssystemen.
\newblock Springer (1993)  27--48

\bibitem{bernardi2012dependability}
Bernardi, S., Merseguer, J., Petriu, D.C.:
\newblock {Dependability Modeling and Analysis of Software Systems Specified
  with UML}.
\newblock ACM Computing Surveys \textbf{45}(1) (2012)  1--48

\bibitem{venkatesan2013survey}
Venkatesan, L., Shanmugavel, S., Subramaniam, C.,  et~al.:
\newblock {A Survey on Modeling and Enhancing Reliability of Wireless Sensor
  Network}.
\newblock Wireless Sensor Network \textbf{5}(03) (2013)  41--51

\bibitem{trivedi2008probability}
Trivedi, K.S.:
\newblock {Probability \& Statistics with Reliability, Queuing and Computer
  Science Applications}.
\newblock John Wiley \& Sons (2008)

\bibitem{fugua2003applicability}
Fugua, N.:
\newblock {The Applicability of Markov Analysis Methods to Reliability,
  Maintainability, and Safety}.
\newblock Reliability Analysis Center START Sheet \textbf{10}(2) (2003)  1--8

\bibitem{distefano2006new}
Distefano, S., Xing, L.:
\newblock {A New Approach to Modeling the System Reliability: Dynamic
  Reliability Block Diagrams}.
\newblock In: Reliability and Maintainability Symposium, IEEE (2006)  189--195

\bibitem{peterson1981Petri}
Peterson, J.L.:
\newblock {Petri Net Theory and the Modeling of Systems}.
\newblock Prentice Hall (1981)

\bibitem{zhong2006petri}
Zhong, D., Qi, Z.:
\newblock {A Petri Net based approach for Reliability Prediction of Web
  Services}.
\newblock In: On the Move to Meaningful Internet Systems. Volume 4277 of LNCS.
\newblock (2006)  116--125

\bibitem{yang2011petri}
Yang, X., Li, J., Liu, W., Guo, P.:
\newblock {Petri Net Model and Reliability Evaluation for Wind Turbine
  Hydraulic Variable Pitch Systems}.
\newblock Energies \textbf{4}(6) (2011)  978--997

\bibitem{kumar2012reliability}
Kumar, G., Jain, V., Gandhi, O.:
\newblock {Reliability and Availability Analysis of Mechanical Systems using
  Stochastic Petri Net Modeling based on Decomposition Approach}.
\newblock International Journal of Reliability, Quality and Safety Engineering
  \textbf{19}(01) (2012)  1--39

\bibitem{jian2008petri}
Jian, S., Shaoping, W., Yaoxing, S.:
\newblock {Petri-nets based Availability Model of Fault-tolerant Server
  System}.
\newblock In: Robotics, Automation and Mechatronics, IEEE (2008)  444--449

\bibitem{dugan1989stochastic}
Dugan, J.B., Ciardo, G.:
\newblock {Stochastic Petri Net Analysis of a Replicated File System}.
\newblock Software Engineering \textbf{15}(4) (1989)  394--401

\bibitem{zengkai2013availability}
Zengkai, L., Yonghong, L., Ju, L.:
\newblock {Availability and Reliability Analysis of Subsea Annular Blowout
  Preventer}.
\newblock In: International Conference on Energy. Volume~25., Science \&
  Engineering Research Support Society (2013)  73--76

\bibitem{beirong2012availability}
Beirong, Z., Xiaowen, X., Wei, X.:
\newblock {Availability Modeling and Analysis of Equipment based on Generalized
  Stochastic Petri Nets}.
\newblock Research Journal of Applied Sciences, Engineering and Technology
  \textbf{4}(21) (2012)  4362--4366

\bibitem{snoopy}
Heiner, M., Herajy, M., Liu, F., Rohr, C., Schwarick, M.:
\newblock {SNOOPY - A Unifying Petri Net Tool}.
\newblock In: Application and Theory of Petri Nets. Volume 7347 of LNCS.
\newblock Springer (2012)  398--407

\bibitem{beaudouin2001cpn}
Beaudouin-Lafon, M.,  et~al.:
\newblock {CPN/Tools: A Tool for Editing and Simulating Coloured Petri Nets}.
\newblock In: Tools and Algorithms for the Construction and Analysis of
  Systems. Volume 2031 of LNCS.
\newblock Springer (2001)  574--577

\bibitem{wei2011dependability}
Wei, B., Lin, C., Kong, X.:
\newblock {Dependability Modeling and Analysis for the Virtual Data Center of
  Cloud Computing}.
\newblock In: High Performance Computing and Communications, IEEE (2011)
  784--789

\bibitem{guimaraes2011dependability}
Guimar{\~a}es, A., Maciel, P., Matos~Jr, R., Camboim, K.:
\newblock {Dependability Analysis in Redundant Communication Networks using
  Reliability Importance}.
\newblock In: Information and Network Technology. Volume~4., IACSIT Press
  (2011)  12--17

\bibitem{omidi2012modeling}
Omidi, A., Moradi, S.:
\newblock {Modeling and Quantitative Evaluation of an Internet Voting System
  based on Dependable Web Services}.
\newblock In: Computer and Communication Engineering, IEEE (2012)  825--829

\bibitem{lijie2012verification}
Lijie, C., Tao, T., Xianqiong, Z., Schnieder, E.:
\newblock {Verification of the safety communication protocol in train control
  system using colored Petri net}.
\newblock Reliability Engineering \& System Safety \textbf{100} (2012)  8--18

\bibitem{li2014calculation}
Li, Y.z., Yi, H.y.:
\newblock {Calculation Method on Reliability of Logistics Service Supply Chain
  Based on Stochastic Petri Nets}.
\newblock International Journal of u-and e-Service, Science and Technology
  \textbf{7}(1) (2014)  103--112

\bibitem{robidoux2010automated}
Robidoux, R., Xu, H., Xing, L., Zhou, M.:
\newblock {Automated Modeling of Dynamic Reliability Block Diagrams using
  Colored Petri Nets}.
\newblock Systems, Man and Cybernetics, Part A: Systems and Humans
  \textbf{40}(2) (2010)  337--351

\bibitem{nebel2008modeling}
Nebel, S., Bertsche, B.:
\newblock {Modeling and Simulation Methodology of the Operational Availability
  and Logistics using Extended Colored Stochastic Petri Nets—an Astronautics
  Case Study}.
\newblock In: Reliability and Maintainability Symposium, IEEE (2008)  434--439

\bibitem{sadou2009reliability}
Sadou, N., Demmou, H.:
\newblock {Reliability Analysis of Discrete Event Dynamic Systems with Petri
  Nets}.
\newblock Reliability Engineering \& System Safety \textbf{94}(11) (2009)
  1848--1861

\bibitem{balakrishnan1996stochastic}
Balakrishnan, M., Trivedi, K.S.:
\newblock {Stochastic Petri Nets for the Reliability Analysis of Communication
  Network Applications with Alternate-routing}.
\newblock Reliability Engineering \& System Safety \textbf{52}(3) (1996)
  243--259

\bibitem{radevanalysis}
Radev, D., Rashkova, E., Denchev, V.:
\newblock {Analysis of Markov Reward Models with Stochastic Petri Nets}.
\newblock In: International Conference on Computer Systems and Technologies,
  ACM (2008)  1--6

\bibitem{kohlik2009dependability}
Kohl{\'\i}k, M.:
\newblock {Dependability Models based on Petri Nets and Markov Chains} (2009)

\bibitem{zhu2012service}
Zhu, L., Yu, F.R., Ning, B., Tang, T.:
\newblock {Service Availability Analysis in Communication-based Train Control
  systems using WLANs}.
\newblock In: Communications, IEEE (2012)  1383--1387

\bibitem{jindal2011markov}
Jindal, V., Dharmaraja, S., Trivedi, K.S.:
\newblock {Markov Modeling Approach for Survivability Analysis of Cellular
  Networks}.
\newblock International Journal of Performability Engineering \textbf{7}(5)
  (2011)  429

\bibitem{schoenen2013erlang}
Schoenen, R., Yanikomeroglu, H.:
\newblock {Erlang Analysis of Cellular Networks using Stochastic Petri Nets and
  User-in-the-loop Extension for Demand Control}.
\newblock In: Global Communication Conference, IEEE (2013)  298--303

\bibitem{youness2006robust}
Youness, O., Elkilani, W., El-Wahed, W.A., Torkey, F.:
\newblock {A Robust Methodology for Performance Evaluation of Communication
  Networks Protocols}.
\newblock In: Communication Networks and Services Research Conference, IEEE
  (2006)  1--10

\bibitem{christodoulou1994petri}
Christodoulou, S., Zhou, M.:
\newblock {A Petri Net Approach to Modeling and Performance Analysis of Fiber
  Data Distributed Interface (FDDI) Network}.
\newblock In: Emerging Technologies and Factory Automation, IEEE (1994)
  373--380

\bibitem{ibe1993performance}
Ibe, O.C., Choi, H., Trivedi, K.S.:
\newblock {Performance Evaluation of Client-server Systems}.
\newblock Parallel and Distributed Systems \textbf{4}(11) (1993)  1217--1229

\bibitem{279721}
Tunik, A., Kharlashkin, I.:
\newblock {A Formalistic Method for the Performance Evaluation of Communication
  Networks of Distributed Computing Systems}.
\newblock In: Industrial Electronics. Volume~2., IEEE (1992)  874--878

\bibitem{sun2009survivability}
Sun, X., Lin, C., Liu, W., Xiao, Y.:
\newblock {Survivability Evaluation of Distributed Service using Stochastic
  Petri Net}.
\newblock In: Communications and Networking in China, IEEE (2009)  1--5

\bibitem{zeng2011spn}
Zeng, W., Hong, Z.G.:
\newblock {SPN-based Performance Analysis of LEO Satellite Networks with
  Multiple Users}.
\newblock In: Machine Learning and Cybernetics. Volume~3., IEEE (2011)
  1425--1429

\bibitem{choi1994markov}
Choi, H., Kulkarni, V.G., Trivedi, K.S.:
\newblock {Markov Regenerative Stochastic Petri Nets}.
\newblock Performance Evaluation \textbf{20}(1) (1994)  337--357

\bibitem{baier2008principles}
Baier, C., Katoen, J.P.:
\newblock {Principles of Model Checking}.
\newblock MIT Press (2008)

\bibitem{lin2010modeling}
Lin, C.M., Yang, C.W., Teng, H.K., Chung, M.C., Lang, K.C., Teng, H.F.:
\newblock {Modeling CAN Network using PRISM}.
\newblock In: Industrial Informatics, IEEE (2010)  390--394

\bibitem{hermanns2003etmcc}
Hermanns, H., Katoen, J.P., Meyer-Kayser, J., Siegle, M.:
\newblock {ETMCC: Model Checking Performability Properties of Markov Chains}.
\newblock In: Dependable Systems and Networking, IEEE (2003) ~1

\bibitem{pervez2014formal}
Pervez, U., Hasan, O., Latif, K., Tahar, S., Gawanmeh, A., Hamdi, M.S.:
\newblock {Formal Reliability Analysis of a Typical FHIR Standard based
  e-Health System using PRISM}.
\newblock In: e-Health Networking, Applications and Services, IEEE (2014)
  43--48

\bibitem{pervez2015formal}
Pervez, U., Mahmood, A., Hasan, O., Latif, K., Gawanmeh, A.:
\newblock {Formal Reliability analysis of Device Interoperability Middleware
  (DIM) based E-health system using PRISM}.
\newblock In: e-Health Networking, Applications and Services. (2015)  1--6

\bibitem{gomes2010systematic}
Gomes, A., Mota, A., Sampaio, A., Ferri, F., Buzzi, J.:
\newblock {Systematic Model-based Safety Assessment via Probabilistic Model
  Checking}.
\newblock In: Leveraging Applications of Formal Methods, Verification, and
  Validation. Volume 6415 of LNCS.
\newblock Springer (2010)  625--639

\bibitem{gopinath2009reliability}
Gopinath, K., Elerath, J., Long, D.:
\newblock {Reliability Modelling of Disk subsystems with Probabilistic Model
  Checking}.
\newblock Technical report, Technical Report UCSC-SSRC-09-05, University of
  California, Santa Cruz (2009)
  \url{http://www.crss.ucsc.edu/media/papers/ssrctr-09-05.pdf}.

\bibitem{ge2010analysing}
Ge, X., Paige, R.F., McDermid, J.A.:
\newblock {Analysing System Failure Behaviours with PRISM}.
\newblock In: Secure Software Integration and Reliability Improvement
  Companion, IEEE (2010)  130--136

\bibitem{peng2013probabilistic}
Peng, Z., Lu, Y., Miller, A., Johnson, C., Zhao, T.:
\newblock {A Probabilistic Model Checking Approach to Analysing Reliability,
  Availability, and Maintainability of a Single Satellite System}.
\newblock In: Modelling Symposium, IEEE (2013)  611--616

\bibitem{bozzano2009compass}
Bozzano, M., Cimatti, A., Katoen, J.P., Nguyen, V.Y., Noll, T., Roveri, M.:
\newblock {The {COMPASS} Approach: Correctness, Modelling and Performability of
  Aerospace Systems}.
\newblock In: Computer Safety, Reliability, and Security. Volume 5775 of LNCS.
\newblock Springer (2009)  173--186

\bibitem{cimatti2002nusmv}
Cimatti, A., Clarke, E., Giunchiglia, E., Giunchiglia, F., Pistore, M., Roveri,
  M., Sebastiani, R., Tacchella, A.:
\newblock {NuSMV 2: An Opensource Tool for Symbolic Model Checking}.
\newblock In: Computer Aided Verification. Volume 2404 of LNCS.
\newblock (2002)  359--364

\bibitem{katoen2005Markov}
Katoen, J.P., Khattri, M., Zapreev, I.S.:
\newblock A {M}arkov {R}eward {M}odel {C}hecker.
\newblock In: Quantitative Evaluation of Systems, IEEE (2005)  243--244

\bibitem{norman2005evaluating}
Norman, G., Parker, D., Kwiatkowska, M., Shukla, S.:
\newblock {Evaluating the Reliability of NAND Multiplexing with PRISM}.
\newblock Computer-Aided Design of Integrated Circuits and Systems
  \textbf{24}(10) (2005)  1629--1637

\bibitem{norman2014quantitative}
Norman, G., Parker, D.:
\newblock Quantitative {V}erification: {F}ormal {G}uarantees for {T}imeliness,
  {R}eliability and {P}erformance.
\newblock Technical report (2014)

\bibitem{conghua2013analysis}
Conghua, Z., Meiling, C.:
\newblock {Analysis of Fast and Secure Protocol based on Continuous-time Markov
  Chain}.
\newblock Communications, China \textbf{10}(8) (2013)  137--149

\bibitem{hurd_02}
Hurd, J.:
\newblock Formal {V}erification of {P}robabilistic {A}lgorithms.
\newblock Ph{D} {T}hesis, University of Cambridge, UK (2002)

\bibitem{mhamdi_lebsegue}
Mhamdi, T., Hasan, O., Tahar, S.:
\newblock On the {F}ormalization of the {L}ebesgue {I}ntegration {T}heory in
  {HOL}.
\newblock In: Interactive {T}heorem {P}roving. Volume 6172 of LNCS.
\newblock Springer (2010)  387--402

\bibitem{holzl2011three}
H{\"o}lzl, J., Heller, A.:
\newblock {Three Chapters of Measure Theory in Isabelle/HOL}.
\newblock In: Interactive Theorem Proving. Volume 6898 of LNCS.
\newblock Springer (2011)  135--151

\bibitem{hasan2011formal}
Hasan, O., Patel, J., Tahar, S.:
\newblock {Formal Reliability Analysis of Combinational Circuits using Theorem
  Proving}.
\newblock Journal of Applied Logic \textbf{9}(1) (2011)  41--60

\bibitem{hasan2010formal}
Hasan, O., Tahar, S., Abbasi, N.:
\newblock {Formal Reliability Analysis using Theorem Proving}.
\newblock Transactions on Computers \textbf{59}(5) (2010)  579--592

\bibitem{abbasi2010formal}
Abbasi, N., Hasan, O., Tahar, S.:
\newblock {Formal Lifetime Reliability Analysis using Continuous Random
  Variables}.
\newblock In: Logic, Language, Information and Computation. Volume 6188 of
  LNCS.
\newblock Springer (2010)  84--97

\bibitem{WAhmad_CICM14}
Ahmed, W., Hasan, O., Tahar, S., Hamdi, M.S.:
\newblock {Towards the Formal Reliability Analysis of Oil and Gas Pipelines}.
\newblock In: Conferences on Intelligent Computer Mathematics. Volume 8543 of
  LNCS.
\newblock Springer (2014)  30--44

\bibitem{WAhmed_Wimob15}
Ahmed, W., Hasan, O., Tahar, S.:
\newblock {Formal Reliability Analysis of Wireless Sensor Network Data
  Transport Protocols using HOL}.
\newblock In: Wireless and Mobile Computing, Networking and Communications,
  IEEE (2015)  217--224

\bibitem{WAhmed_IWIL15}
Ahmed, W., Hasan, O., Tahar, S.:
\newblock {Towards Formal Reliability Analysis of Logistics Service Supply
  Chains using Theorem Proving}.
\newblock In: Implementation of Logics. (2015)  111--121

\bibitem{WAhmad_CICM15}
Ahmed, W., Hasan, O.:
\newblock {Towards Formal Fault Tree Analysis Using Theorem Proving}.
\newblock In: Intelligent Computer Mathematics. Volume 9150 of LNCS.
\newblock Springer (2015)  39--54

\bibitem{liu2013formal}
Liu, L., Hasan, O., Tahar, S.:
\newblock {Formal Reasoning About Finite-State Discrete-Time Markov Chains in
  {HOL}}.
\newblock Journal of Computer Science and Technology \textbf{28}(2) (2013)
  217--231

\bibitem{holzl2012interactive}
H{\"o}lzl, J., Nipkow, T.:
\newblock {Interactive Verification of Markov Chains: Two Distributed Protocol
  Case Studies}.
\newblock arXiv preprint arXiv:1212.3870 (2012)

\bibitem{slind2008brief}
Slind, K., Norrish, M.:
\newblock {A Brief Overview of HOL4}.
\newblock In: Theorem Proving in Higher Order Logics. Volume 5170 of LNCS.
\newblock Springer (2008)  28--32

\end{thebibliography}

\end{document}